\documentstyle[12pt]{article}
\begin{document}
\begin{center}
{\bf TOWARDS A NON-PERTURBATIVE
 RENORMALIZATION OF EUCLIDEAN
 QUANTUM GRAVITY}\\
\vspace{1cm}
Z. Burda$^a$\footnote{Permanent address: 
Institute of Physics, 
Jagellonian University, ul. Reymonta 4, PL-30 059,
Krak\'{o}w, Poland }, J.-P. Kownacki$^b$ 
and A. Krzywicki$^b$\\
\vspace{0.7cm}
$^a$ Fakult\"{a}t f\"{u}r Physik,
 Universit\"{a}t Bielefeld, 
Bielefeld 33501, Germany\\
$^b$ LPTHE,
B\^{a}t 211, Universit\'{e} 
de Paris-Sud, 91405 Orsay,
 France\footnote{Laboratoire 
associ\'{e} au CNRS, URA-D0063}\\
\end{center}
\vspace{2cm}
{\bf Abstract}: A real space
renormalization group technique, 
based on the hierarchical baby-universe
structure of a typical dynamically 
triangulated manifold, is used
to study scaling properties of 
2d and 4d lattice 
quantum gravity. In 4d, the
$\beta$-function is defined and
calculated numerically.  An
evidence for the existence of an 
ultraviolet stable fixed point of the 
theory is presented.
\par\noindent

\vspace{2cm}
\noindent
May 1995\\
LPTHE Orsay 95/34\\
\newpage
{\bf 1.} As is well known, the
 renormalization group (RG) 
is a tool providing deep insight into the
structure of a quantum field theory.
 It is certainly worth applying
this tool to quantum gravity. This 
work is devoted to the 
development and the application
of the real space renormalization group technique
in the context of euclidean 
quantum gravity. It is a direct
continuation of the work \cite{jkk}. 
\par
We choose the lattice gravity
 framework. More 
precisely we adopt the 
particularly promising 
dynamical triangulation 
approach \cite{dt}.
The remarkable results
obtained within a class
of exactly solvable models 
in two dimensions strongly suggest
that the dynamical triangulation 
recipe is the correct way of 
discretizing gravity (at least
for fixed topology). 
\par
In conventional statistical
mechanics a real space
renormalization group transformation 
has two facets:
\par
(a) {\em geometry} - cells of the body
 are "blocked" together.
\par
(b) {\em matter fields} - "block" fields
 are defined in 
terms of the original fields.
\par
On a regular lattice it is trivial
 to perform the step
(a) in such a manner that the 
resulting lattice is identical,
modulo
rescaling, to the original one.
 Since the values of
critical couplings depend on 
the lattice type this
self-similarity feature of the 
transformation is important.
On a random lattice an appropriate
definition of (a) requires
some thought. In this work we 
consider pure geometry, without
matter fields, and consequently
 the geometrical
aspect of the renormalization group.
\par
In ref. \cite{jkk} a method of 
"blocking" triangulations that
exploits the self-similarity
feature of random manifolds has
been proposed. Without repeating
in detail the arguments
presented in \cite{jkk} let us 
briefly sketch the main idea~\footnote{As 
we have learned from J.
Ambj\o rn during the LATTICE 
'94 Conference, our ideas partly 
overlap with those discussed
earlier for 2d in ref. \cite{amb1}.} .
The intuitive arguments given
below will be replaced progressively
by more precise ones
later on. We do not wish to give
an impression of complexity
from the outset. 
\par
{\bf 2.} In 2d one can show \cite{jm} that an 
infinite randomly triangulated manifold 
is a self-similar tree obtained
by gluing together 
sub-structures called baby
universes (BUs), defined
as subuniverses
separated from the remaining part
of the universe by a narrow neck.
We conjecture
that a similar picture holds
in 4d, at least in the neighborhood
of the phase transition point. The results
presented in this paper strongly
support this conjecture. 
\par
We have proposed \cite{jkk} to define the
step (a) as the operation of
cutting the last generation of BUs,
i.e.  those which have no
further BUs growing on them.
When the tree is finite, and as
one continues cutting the successive 
generations of BUs, it
gets smaller (in lattice units) 
and looks less and less branched.
In the spirit of the renormalization 
group this is interpreted as 
reflecting the reduction of the
resolving power: although, in physical 
units, the manifold remains of
the same size,  the fine 
structure is not observed.
\par
The next problem consists in 
establishing a connection between
lattice and physical observables.
 For example, in 
lattice QCD one identifies the 
inverse of the correlation length   
$\xi$ with the mass $m$ of a hadron
\begin{equation}
m a = \xi^{-1}
\label{1}
\end{equation}
\noindent
Here $a$ denotes the lattice 
spacing and $\xi =\xi(\kappa)$
depends on a dimensionless coupling
constant $\kappa$.
The correlation length $\xi$ can 
be calculated for a given
$\kappa$. The mass $m$ is a 
physical observable, independent
of the regularization procedure.
 Hence, eq. (\ref{1}) yields
the variation of the lattice 
spacing $a$, measured in physical
units, with the coupling $\kappa$.
 The so-called $\beta$-function
can by defined as
\begin{equation}
\beta(\kappa) = {{d\kappa} 
\over {d \ln (1/a)}} \; ,
\label{2}
\end{equation}
where the derivative  
is taken along the RG flow.
In QCD, the lattice correlation 
length diverges exponentially
as $\kappa \to 0$ : $\xi \propto
 \exp (1/2\beta_0 \kappa^2)$
with $\beta_0 > 0$. Eq. (\ref{1})
 implies that
$\beta(\kappa) = - \beta_0 \kappa^3$.
 The fixed point is at
$\kappa = 0$ and is ultraviolet 
(UV) stable. The lattice spacing
tends to zero, in physical units 
set by eq. (\ref{1}), as one
approaches the fixed point.
\par
We recall these very well known
 things, in a sense, to set the
standards. We would like to have
 similar arguments in quantum
gravity. The phase transition 
observed in 4d \cite{aj,am} seems
to be continuous and therefore 
one has a candidate for a fixed
point. The nature of this fixed 
point (IR or UV) is to be found.
The soundness of the proposed 
renormalization group has to be
further examined. 
Finally, one has to select
appropriate observables in order
 to fix the scales. We shall
report about some progress 
achieved in studying these problems.
But before proceeding we have 
to define the theory we are working with.
\par
{\bf 4.}  We take an euclidean 
version of the Einstein--Hilbert
action which for a simplicial manifold reads
\begin{equation}
S = - \kappa_2 N_2 + \kappa_4 N_4 \; \; ,
\label{3}
\end{equation}
where $N_2$ and $N_4$ denote 
the total number of triangles and
 4-simplexes, respectively. 
The theory is
defined by the partition function
\begin{equation}
Z(\kappa_2, \kappa_4) = 
\sum_{N_2,N_4} Z_{N_2 N_4} e^{-S}
\label{4}
\end{equation}
\noindent
where 
\begin{equation}
Z_{N_2 N_4} = \sum_{T(N_2,N_4)} W(T)
\label{5}
\end{equation}
The sum in eq. (\ref{5}) runs 
over fixed topology 4d simplicial manifolds 
$T(N_2,N_4)$ with 
$N_2$ triangles and $N_4$ 4-simplexes. 
$W(T)$ is the symmetry factor taking 
care of the equivalent relabelings
of the manifold.
\par
We study a fixed volume 
$N_4=N$ canonical ensemble for
spherical manifolds with 
the partition function~:
\begin{equation}
Z(\kappa_2, N_4) =
 \sum_{N_2} Z_{N_2 N_4} e^{-S}
\label{6}
\end{equation}
In practice,
in order to have an ergodic algorithm, 
one is forced to simulate
 a multicanonical ensemble 
(see e.g. \cite{bbj}). One 
lets the volume fluctuate
in an external potential $U(N_4)$ with 
parameters adjusted in such a 
way that the entry $N_4=N$, at
which measurements are done, 
is highly probable in 
the resulting $N_4$ distribution.
In our simulations we  have
used $U = {{\delta} \over 2} (N-N_4)^2$
with $\delta = 0.001$.
\par
The observables have to be 
geometrical objects. One obvious choice
is the total physical volume 
$V = N a^4$. Once $V$ has 
been fixed, the
length of the lattice step 
in physical units is also 
fixed for a given
lattice:

\begin{equation}
a = (V/N)^{1 \over 4}
\label{7}
\end{equation}

\noindent
However, the continuum limit
 of the theory is not yet defined.
A second observable is needed 
for that. A priori, this second observable
could be the local curvature 
$R$. It is, however, not quite obvious
how the {\em physical curvature}  
should actually be defined. In numerical
simulations of dynamically 
triangulated 4d manifolds, the bare Regge
curvature residing on the 
hinges of the lattice is found to be
nonvanishing at the transition
point \cite {aj,am,amb2}, which leads to basic
problems in defining its 
dimensionful counterpart in
the continuum limit.
Various interpretations have 
been given to this finding. We are
tempted to adopt here the 
point of view formulated by the 
authors of ref. \cite{dbs},
who argue in substance that
the bare curvature is a lattice
artifact and is not the 
curvature that would measure an observer
using a physical stick,
independent of the cut-off.
\par
Following the suggestion 
put forward in \cite{dbs},
we choose as our second observable a
length scale characterizing 
the large-distance behavior of the
so-called puncture-puncture 
correlations function, viz. the average
geodesic distance between two randomly
selected simplexes of the manifold:
\begin{equation}
\langle r \rangle = \langle N^{-2}
 \sum_{ab} r_{ab} \rangle_N  \; \; ,
\label{8}
\end{equation}
\noindent
where $r_{ab}$ is the geodesic
distance between 4-simplexes $a$ and $b$
and $\langle ... \rangle_N$ denotes the
average over manifolds described by the
canonical partition function (\ref{6}).
\par
In practice, it is difficult to
identify BUs with necks larger than
the minimum one. Following the 
terminology introduced in ref. \cite{jm}
we call minBUs the baby universes 
with the smallest possible neck.
We have sugested in \cite{jkk} 
that a meaningful RG
transformation is defined by
 the operation of cutting last
generation minBUs. Using the 
pure 2d quantum gravity as a toy model,
we have checked that, on the 
average, the area of surfaces obtained
after cutting last generation 
minBUs is, for large surfaces, a
constant fraction of the area 
of the original surface. It is
important to check that a similar
 result holds for $\langle r \rangle$,
if $\langle r \rangle$ is to 
be used in a RG analysis.
\par
{\bf 5.} We have been working 
with surfaces of area
ranging between $N = 4096$
and 65536 triangles. The minimum
number of triangles in a minBU has
been set to 15. The number
of heating sweeps between measures
has been set to twice the 
autocorrelation length determined, for
the observable $\langle r \rangle$,
in ref. \cite{amb3} \footnote{Strictly speaking,
they have done this for a model
with 2d gravity coupled to a gaussian
field; it does not seem that 
autocorrelation lengths are longer for
pure gravity when one measures
such a purely geometrical object as the
average geodesic distance.}. The
results are shown in Table 1. The
subscripts ``in'' and ``out'' 
refer to the original and the cut
surfaces, respectively. Actually,
the size of the original surface has  
been kept fixed : $ N_{in} \equiv N$.
\par
It appears from the results 
presented in the Table 1 that the
ratios $\lambda_N = \langle 
N_{out} \rangle  /  N_{in}$ 
and  $\lambda_r = \langle r_{out}
 \rangle  / \langle r_{in} \rangle$
converge towards limiting values
 when $N \to \infty$. Furthermore,
the critical exponent $\nu$ defined
 by the relation
$\lambda_r =\lambda_N^{\nu}$
 approaches the expected 
value ${1 \over 4}$.
Let us recall that the problem
 of scaling in pure 2d gravity has been
solved rigorously \cite{kaw} 
and it is known that
$\langle r  \rangle \propto 
N^{1 \over 4}$ as $N \to \infty$. This
could have been checked numerically
 with better precision using
a more direct method (see e.g. 
ref. \cite{bow}). Our point is that 
$\langle r  \rangle$ scales 
correctly under the RG
transformation we have proposed.
 Notice, that self-similarity and
scaling only hold in the 
statistical sense: it is the grand-canonical 
{\em ensemble} that is 
mapped into itself by our transformation.
\par
In 2d gravity one can write for small enough $a$ 

\begin{equation}
\langle r  \rangle = (\rho/a)^{2\nu} 
\; \; (\nu = {1 \over 4})
\label{11}
\end{equation}

\noindent
where $\rho$ is an observable with 
the dimension of a length,
which indirectly determines the
 physical curvature of the surface. 
Eq. (\ref{11}) is an analogue of
 eq. (\ref{1}) and $\rho$ could be
used to set the scale of the 
theory. However, once the physical area
$V = Na^2$ has been used for 
that purpose, the choice of
$\rho$ is no longer free. 
\par
Playing with the toy model we
 have convinced ourselves that
$\langle r \rangle$ is an observable 
suitable for our purposes.
We also got better acquainted with 
the proposed RG
transformation. Thus, when one 
cuts the successive
generations of minBUs, the 
corresponding scaling factors
$\lambda_N^{(1)}$, 
 $\lambda_N^{(2)}$, ... , 
 $\lambda_N^{(n)}$, ...  
are not at all equal and actually
 rapidly tend to unity. This
is easily understood from the
 results of ref. \cite{jm}: before
the outer minBUs have been cut,
 the minBUs of the next layer
are obviously more volumineous 
than the outer ones (since the latter 
are just outgrows off the former).
 In order to keep $\lambda_N$
constant, one should enlarge 
the necks of BUs to be cut at each
iteration of the RG 
transformation. This
is not what we are doing,
 since we restrict ourselves,
for purely technical reasons, 
to minBUs. 
\par
The results shown in Table 1
 have been obtained iterating
the transformation until no 
minBU is left. When only one
iteration is performed, the 
value of $\lambda_N$ is
larger (about 0.86 , with the
 same definition of a minBU)
and $\lambda_r$ is very close
 to unity. Consequently,
the error in the determination 
of $\nu$ is much
larger, for the same statistics,
 and the results appear
less convincing.
\par
{\bf  6.} Encouraged by the 
results obtained in 2d 
we go over to 4d. 
Now, $\langle r  \rangle$ 
is a function of two variables, viz. $N$ 
and $\kappa_2$, to be denoted
 $\kappa$ from here on for the sake
of simplicity of writing.
\par
Suppose we start with $N = N_{in}$ 
and $\kappa = \kappa_{in}$
and calculate the corresponding 
average geodesic distance
$\langle r  \rangle = \langle 
r_{in}  \rangle$. After
one iteration of RG we find
$\langle N_{out}  \rangle$ and 
$\langle r_{out}  \rangle$.
Denote by $\kappa_{out}$ 
the value of $\kappa$  characterizing
the new ensemble. Define

\begin{eqnarray}
\delta r & = & \langle r_{in} 
 \rangle - \langle r_{out}  \rangle \\
\delta N & = & N_{in} - 
\langle N_{out}  \rangle \\
\delta \kappa & = & \kappa_{in}
 - \kappa_{out} 
\label{12}
\end{eqnarray}

\noindent From (\ref{7}) we get

\begin{equation}
\delta \ln (1/a) = {1 \over 4} 
\delta \ln N
\label{13}
\end{equation}

\noindent
The associated shift of $\kappa$ 
can be found from
the equation

\begin{equation}
\delta r = r_N \delta \ln N +
 r_{\kappa} \delta \kappa
\label{14}
\end{equation}

\noindent
where $r_N$ and $r_{\kappa}$ 
are the partial derivatives
of $\langle r  \rangle$ with
 respect to $\ln N$ and $\kappa$,
which can also be estimated 
from numerical simulations. Once
$ \delta \kappa$ has been 
calculated, the $\beta$-function is
readily found from (\ref{2}) 
and (\ref{13}). Writing eq. (15) we 
have actually conjectured that $\kappa$
is the only relevant coupling 
at long distances. 
This is presumably reasonable when $\kappa$ 
is close enough to 
its critical value. One can
check the conjecture by calculating 
the $\beta$-function
using some other observable than 
$\langle r \rangle$, to fix
the scale. If the conjecture 
is true the result should not
depend appreciably on the 
choice of the observable.
\par
We have applied our RG 
transformation to 4d spherical 
simplicial complexes with
8000 simplexes, working with a 
series of values of $\kappa$
ranging between 0.9 and 1.45 . 
The average geodesic distance
has also been calculated, for 
the same values of $\kappa$
at $N = 6000$, in order to 
estimate $r_N$. The five standard local moves
have been used for updating. 
Typically, we have
been carrying $\sim 10^5$ sweeps 
of the lattice, for each $\kappa$.
Close to the critical 
point the number of sweeps has been 
significantly larger. Measures
 were separated by a number
of heating sweeps, equal to 
twice the autocorrelation
length, estimated in advance.
 The errors were found using
the conventional binning method.
 The minimum volume of a 
minBU has been set to 20 
simplexes.  The RG 
transformation has been iterated 
only once. Above $\kappa = 1.15$  
cutting more than one generation of minBUs 
 yields too 
small $\langle N_{out} \rangle$ 
for the differential 
formula (\ref{14}) to be trusted 
\footnote{At $\kappa = 1.3$ , 
iterating RG the maximum 
number of times one 
finds $\langle N_{out} \rangle 
= 859  \pm 300$: the 
manifolds resemble branched polymers 
and the situation differs from that
described in sect. 5.} 
 The data are collected in
Table 2 and the $\beta$-function
 is shown in Fig. 1.
A linear fit to the data yields 
$\beta(\kappa) = \beta_0 (\kappa^* -
\kappa)$ with $\beta_0 = 6.5(1.2)$ 
and $\kappa^* = 1.2(2)$.
\par
Although the errors are large, 
especially as one enters the 
branched polymer phase, where 
the local moves are not very 
efficient, it is rather evident
from Fig. 1 that the critical point is an
ultraviolet stable one\footnote{P. Bialas
has informed us that using the 
data of ref \cite{dbs}
and our definition of the 
$\beta$-function one 
can estimate the latter in
3 points. The estimates 
agree with our result.}. 
Notice, that a 
qualitatively identical 
result has 
been found analytically 
in $2+ \epsilon$ dimensions
\cite{wein,kaw2}, using 
perturbation theory together with
the $\epsilon$ expansion.
 Let us now briefly
outline some general consequences of
this result.
\par
{\bf 7.} Close enough to the
 fixed point $\kappa = \kappa^*$ one can
write

\begin{equation}
{{d \kappa} \over {d \ln (1/a)}} =
 \beta_0 (\kappa^* - \kappa) 
\; \; (\beta_0 > 0)
\label{15}
\end{equation}

\noindent
Solving this differential 
equation one finds

\begin{equation}
a = a_0 \mid \kappa^* - 
\kappa \mid^{1/\beta_0} \; \; (\beta_0 > 0)
\label{16}
\end{equation}

\noindent
where $a_0$ is an integration
 constant, analogous to
$\Lambda^{-1}$ in QCD, and 
which should be given a value,
in physical units, in order 
to define the theory. Comparing with
(\ref{7}) one gets

\begin{equation}
V/a_0^4 = t
\label{17}
\end{equation}

\noindent
with $t$ defined by

\begin{equation}
t = N \mid \kappa^* -\kappa 
\mid^{4/\beta_0} 
\label{18}
\end{equation}

The quantity on the left-hand 
side of (\ref{17}) is a constant,
determined by the choice of $V$ 
and $a_0$. The trajectories of the RG flow 
are the curves $t =$  const.
The continuum limit is the double-scaling
limit: $N \to \infty , \; \kappa
 \to \kappa^*$ and $t =$ const.
In analogy to 2d one can write   
$\langle r  \rangle = (\rho/a)^{4\nu}$, 
but now $\rho$ and $\nu$ may 
depend on the value given to $t$.
\par
{\bf 8.} 
Going farther ahead is beyond the
scope of this paper.  
A few remarks are in order at this point~:
\par
(i) This study should and will 
be extended to larger lattices.
In order to do that it will be 
necessary to use a more efficient
updating algorithm. Indeed, the
local algorithm becomes inefficient
as one enters the ``cold'' phase
of the theory, where the
manifold develops a branched 
polymer structure. The ``BU surgery''
algorithm proposed in \cite{amb3,amb4} 
should be appropriate there. 
Performing more intensive
$MC$ studies one should also 
be able to estimate the
corrections to the formula 
(\ref{14}) coming from 
irrelevant directions at the
 fixed point, neglected here.
\par
(ii) Remember, that $\nu^{-1}$ has
 the significance of the
intrinsic Haussdorf dimension $d_H$ of the
 manifold \cite{amb4}. It would be
interesting to look for the
 $t$-dependence of $d_H$ and
determine the class of theories
 satisfying the constraint $d_H = 4$.
Our experience with 2d indicates 
that this may require going to
very large systems.
\par
(iii) It would be very interesting
 to extend the analysis to 
models where matter fields are 
coupled to gravity and to achieve 
a better understanding of the 
relation between the results 
generated by lattice models and
the continuum calculations in 
$2+\epsilon$ dimensions. With 
a more efficient algorithm and
a better statistics it will 
hopefully be possible to examine
the dependence of the shape of 
the $\beta$-function on
the number of matter fields.
Calculations in the continuum
\cite{wein,kaw2} suggest 
something analogous to the 
$c=1$ barrier of 2d models:
the number of matter fields 
cannot be arbitrary for the 
theory to exist. It would
be important to obtain such
a result in lattice gravity,
without any recourse to 
perturbation theory. It might
be that this has some relevance for
the problem of the number of generations.
\par
In summary, we have applied the
previously proposed real space
renormalization group procedure 
to 2d and 4d models of pure euclidean 
quantum gravity. In 4d we have
defined a $\beta$-function and
we have found numerically some 
evidence for the existence of an 
ultraviolet stable fixed point of the 
theory. We have concluded mentioning 
a few problems for the future.
\vspace{1cm}
\par
{\bf Acknowledgements} : We have
benefited from discussions 
and exchange of information
with P. Bialas, D. Johnston,
 J. Jurkiewicz and
B. Petersson. In particular,
J. Jurkiewicz has called our attention
to ref. \cite{dbs} and to
the potential interest of the
distribution of geodesic distances.
The help of B. Klosowicz has 
been invaluable. We are indebted
to the CNRS computing 
center IDRIS,  and 
personally to V. Alessandrini, 
for computer time and cooperation.
One of us (Z.B.) is grateful to
Deutsche Forschungsgemeinschaft
 for financial support.

\vspace{1cm}
\par\noindent
{\bf Figure caption}
\par\noindent
Fig. 1 - The $\beta$-function versus 
the coupling $\kappa$ in 4d,
calculated for spherical 
manifolds with 8000 simplexes.
\newpage

\begin{table}
\centering
\begin{tabular}{|c|c|c|c|c|}
\hline
$N$ & 
$\left<r_{\mbox{\it \tiny 
in}}\right>$ & 
$\frac{\mbox{$\left<{\mbox{\footnotesize 
\it r}}_{\mbox{\it
 \tiny out}}\right>$} }{\mbox{$ 
\left<{\mbox{\it 
\footnotesize r}}_{\mbox{\it 
\tiny in}}\right>$}}\equiv
 \lambda_{r}$ &
 $\frac{\mbox{$\left<\mbox{\footnotesize
 \it N}_{out}
 \right>$}}{\mbox{{\footnotesize 
\it N}}} 
\equiv \lambda_{\mbox{\it
 \tiny N}}$ &
 \mbox{$\vcenter{\vskip 1mm
 \hbox{$\nu$ 
{\rm \footnotesize from}} \vskip 1mm  
 \hbox{$\lambda_{r}=\lambda^{\nu}_{\mbox{\it 
\tiny N}}$} \vskip 1mm  }$} \\
\hline
4096 & 31.20(22) & 0.880(8)
 & 0.745(6) & 0.433(33) \\
8192 & 39.34(24) & 0.900(5) 
& 0.752(7) & 0.371(24) \\
16384 & 48.99(33) & 0.919(8)
 & 0.764(5) & 0.312(37) \\
32768 & 61.29(63) & 0.928(13) 
& 0.763(4) & 0.276(50) \\
65536 & 75.12(98) & 0.935(16) 
& 0.762(3) & 0.247(63) \\
\hline
$\infty$ & $\infty$ & 0.935
 $^{\ (c)}$  & 
0.765 $^{\ (b)}$  & 0.25 
$^{\ (a)}$  \\
\hline
\multicolumn{5}{l}{$^{(a)}$ 
Theoretical result from ref.  
 \cite{kaw}.}\\
\multicolumn{5}{l}{$^{(b)}$ Obtained by linearly
 extrapolating data.}\\
\multicolumn{5}{l}{$^{(c)}$ Obtained from (a) and (b) 
using $\lambda_r = \lambda_N^{\nu}$}\\
\end{tabular}
\caption{Summary of 
data in 2d.}
\label{tab : 2d exponent}
\end{table}

\begin{table}
\centering
\begin{tabular}{|c||c|c|c||c|}
\hline
&\multicolumn{3}{c||}{$N=8000$}
 & $N=6000$ \\
\hline
$\kappa_2$&$\left< r_{in}
 \right>$&$\left< r_{out}
 \right>$&$\left< N_{out} 
\right>$&$\left< r_{in} \right>$ \\
\hline
0.9&11.907(26)&11.538(25) 
&7422(19) & -\\
1.0&12.661(36) & 11.97(10)
 &7013(76) &12.191(10) \\
1.1 & 14.28(14)&13.20(12)&
6586(118)&13.804(41) \\
1.15 & 15.80(21) & 14.58(30)
 & 6464(203)&16.01(33) \\
1.2 & 23.27(95) & 21.67(95)&
6098(143) &22.18(39) \\
1.25 & 31.10(60) & 29.41(60)
&6101(132)&27.89(61) \\
1.3 &34.06(73)&32.35(85) & 
6035(138)&28.93(69) \\
1.4 &36.77(80) & 35.01(85)
 & 5968(160) & 32.05(66) \\
1.45 & 37.55(84) & 35.83(87)
 & 5980(347) & - \\ 
\hline
\end{tabular}
\caption{Summary of data in 4d.}
\label{tab : kappa2 }
\end{table}


\begin{thebibliography}{99}
\bibitem{jkk} D. Johnston,
 J.-P. Kownacki and A. Krzywicki,
in LATTICE '94, Nucl. Phys. 
B (Proc. Suppl.)42 (1995) 728.
\bibitem{dt} D. Weingarten, 
Nucl. Phys. B210 [FS 6] (1982)
 229; F. David, Nucl.Phys. B257
 (1985) 543 ; J. Ambj\o rn, B.
 Durhuus and J. Fr\"{o}hlich, 
Nucl. Phys. B257 (1985) ; V.A. 
Kazakov, I.K. Kostov and A.A. 
Migdal, Phys. Lett. B157 (1985) 295.
\bibitem{amb1} J. Ambj\o rn, B.
 Durhuus, J. Fr\"{o}hlich and 
T. Jonsson, J. Stat. Phys. 55 (1989) 29.
\bibitem{jm} S. Jain and S.D. 
Mathur, Phys. Lett.
 B286 (1992) 239.
\bibitem{aj} J. Ambj\o rn and 
J. Jurkiewicz, Phys. Lett.B278 (1992) 50.
\bibitem{am} M.E. Agishtein and
 A.A. Migdal, Mod. Phys. Lett.
 A7 (1992) 1039; Nucl. Phys. B385 (1992) 395.
\bibitem{bbj} S. Bilke, Z. 
Burda and J. Jurkiewicz,
Comp. Phys. Comm. 85 (1995) 278.
\bibitem{amb2} J. Ambj\o rn, J. 
Jurkiewicz and C.F.Kristjansen,
 Nucl. Phys. B393 (1993) 601.
\bibitem{dbs} B.V. De Bakker
 and J. Smit,Nucl. Phys. B439 (1995) 239.
\bibitem{amb3} J. Ambj\o rn, 
P. Bialas, Z. Burda, J. Jurkiewicz and 
B. Petersson, Phys. Lett. B325 (1995) 337.
\bibitem{kaw} H. Kawai,N. 
Kawamoto, T. Mogami and Y. 
Watabiki, Phys. Lett. B306 (1993) 19 ;
see also J. Ambj\o rn and Y. 
Watabiki, {\em Scaling in 
quantum gravity}, Copenhagen
 preprint NBI-HE-95-01 
(January 1995) and references therein.
\bibitem{bow} S. Catterall, 
G. Thorleifsson, M. Bowick 
and V. John, Syracuse preprint
 SU-4240-607 (April 1995).
\bibitem{wein} S. Weinberg, 
in {\em General
Relativity, an Einstein Centenary 
Survey}, eds. S.W. Hawking and W.
 Israel (Cambridge University 
Press 1979); R. Gastmans, R. 
Kallosh and C. Truffin, Nucl.
 Phys. B133 (1978) 417; S.M.
 Christensen and M.J. Duff,
 Phys. Let. B79 (1978) 213; 
more recently the problem has
 been thoroughly discussed in 
a series of papers of the Japanese 
group \cite{kaw2}.
\bibitem{kaw2} H. Kawai and M. 
Ninomiya, Nucl. Phys. B336 (1990) 
115 ; H. Kawai, Y. Kitazawa and M.
 Ninomiya, Nucl. Phys. B404 (1993) 
280; {\em ibid} B427 (1993) 684 ; 
T. Aida, Y. Kitazawa, H. Kawai and
 M. Ninomiya, Nucl. Phys. B427 
(1994) 158 ; T. Aida, Y. Kitazawa, 
J. Nishimura and A. Tsuchiya, Tokyo
 preprint TIT-HEP-275 (December 1994).
\bibitem{amb4} J. Ambj\o rn and J.
 Jurkiewicz, {\em Scaling in four 
dimensional quantum gravity}, 
Copenhagen preprint NBI-HE-95-05
 (February 1995).
\end{thebibliography}
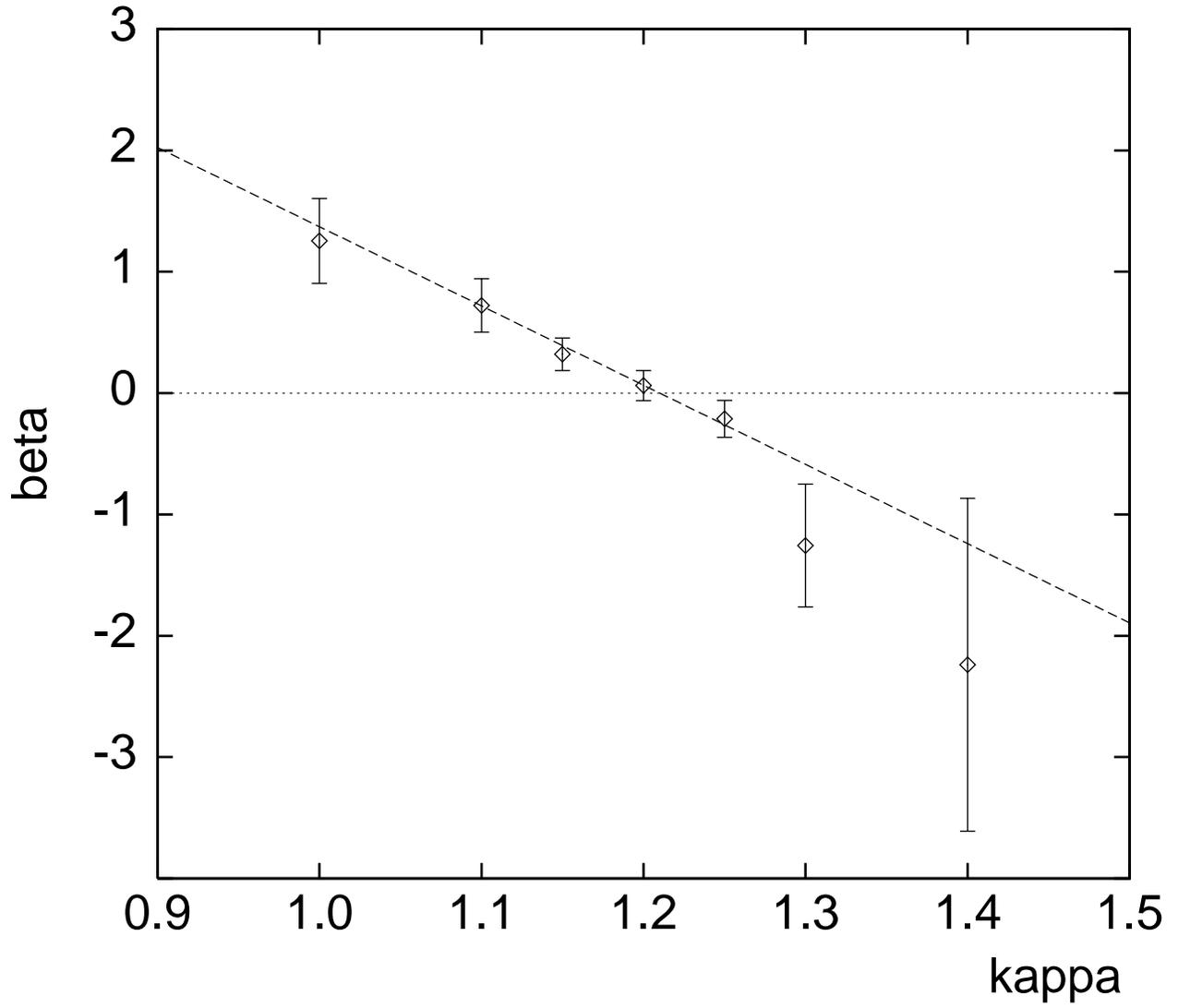
\end{document}